\documentclass{PoS}
\usepackage{bm}

\title{Overview of Spin Physics at EIC}

\ShortTitle{Spin Physics at EIC}

\author{\speaker{Dani\"el Boer}\thanks{I thank my collaborators on the various topics and  Abhay Deshpande, J.H. Lee, and Cristian Pisano for specific feedback and/or discussions.}\\
        Van Swinderen Institute for Particle Physics and Gravity, University of Groningen\\ 
Nijenborgh 4, NL-9747 AG Groningen, The Netherlands\\
        E-mail: \email{d.boer@rug.nl}}

\abstract{The possibilities to measure spin effects at a high-energy Electron-Ion Collider (EIC) are reviewed from a theory point of view. Various types of spin distributions and promising observables are discussed.}

\FullConference{23rd International Spin Physics Symposium - SPIN2018 -\\
		10-14 September, 2018\\
		Ferrara, Italy}

\begin{document}

\section{Introduction}
This overview starts with a discussion of the one-dimensional spin structure of protons and deuterons, given in terms of structure functions and collinear parton distribution functions (PDFs). This is followed by the three-dimensional spin structure (three-dimensional in momentum space), focusing in particular on the Sivers effect for transversely polarized protons and gluon polarization effects in unpolarized protons. Such effects are given in terms of quark and gluon transverse momentum dependent PDFs referred to as TMDs. Here the most promising processes are reviewed. Next fragmentation functions (FFs) are discussed, both collinear and transverse momentum dependent ones, which one can use to do spin physics at the EIC. This focuses mostly on di-hadron FFs and $\Lambda$ FFs. The final section contains a very brief discussion of Generalized TMDs (GTMDs) and Generalized Parton Distributions (GPDs), which capture information about the spatial distribution of partons in the nucleons that can be nontrivially correlated to the transverse momentum and spin.  

\section{1D spin structure}
At an EIC one can first of all perform classic Deep Inelastic Scattering (DIS) measurements, where the objective is to probe the polarized structure functions of which there are five: $g_1$-$g_5$, where $g_3, g_4, g_5$ require weak interactions, hence high $Q^2$. The structure functions $g_1$ and $g_2$ have been measured. The structure function $g_1$ yields information about the proton spin decomposition, through the quark PDFs $\Delta q(x,Q^2)$ and at one order in $\alpha_s$ higher through the gluon PDF $\Delta g(x,Q^2)$. To obtain the contributions of the light quarks and antiquarks to the proton spin $\Delta \Sigma(Q^2)$ one needs to integrate over $x$ and sum over the flavors. Thus far the contribution from the gluons $\Delta g(Q^2) = \int_0^1 dx \Delta g(x,Q^2)$ has been obtained with limited precision \cite{deFlorian:2014yva}, primarily from polarized proton collisions at RHIC. Especially measurements for lower values of $x$ will be of interest here, to reduce the uncertainty from the extrapolation. 

The structure function $g_2$ enters the Burkhardt-Cottingham sum rule ($\int_0^1 dx\ g_2(x,Q^2) = 0$) and has a twist-2 and twist-3 part. The twist-2 part is given in terms of $g_1$. The quantity $d_2 = 3 \int_0^1 dx \left. x^2 \, g_2(x,Q^2)\right|_{\rm twist-3}$ has been extracted from several experiments (see \cite{Armstrong:2018xgk} and references therein) and has been evaluated on the lattice. This $Q^2$-dependent number is found to be small (roughly between $\pm 0.01$), which suggests the relative unimportance of the twist-3 quark-gluon correlation contributions in this case. Likewise $g_3$ has a twist-2 and a twist-3 part. We note however that there is no generally used convention for the definitions of the functions $g_3$-$g_5$ (see \cite{Blumlein:1996vs} for a comparison of definitions). Like the Callan-Gross relation for the unpolarized structure functions, $g_4$ and $g_5$ (in the definition of \cite{Blumlein:1996vs}) are in the parton model related through the Dicus relation $g_4 = 2xg_5$. In charged current DIS the comparison of $g_1$ and $g_5$ can be used to extract information about the charm quark helicity distribution $\Delta c$, see e.g.\ \cite{Zhao:2016rfu}. As said, the measurement of $g_3, g_4, g_5$ requires weak interactions, hence high $Q^2$, which corresponds to a higher $x$ range. The higher the collision energy of the EIC the better for this particular purpose. 
    
Using polarized helium-3 one effectively obtains the structure functions of the polarized neutron, which is interesting in the comparison to the polarized proton (e.g.\ Bjorken sum rule). Alternatively, one can use the deuteron with spectator tagging ($ed \to e'pX$ with $p$ in the target fragmentation region) to access polarized neutron structure \cite{Cosyn:2016oyw}.
Moreover, polarized deuterons allow to probe additional structure functions that are absent for a polarized proton or neutron. In the notation of \cite{Hoodbhoy:1988am} the additional structure functions are $b_1$-$b_4$, where $b_1$ and $b_2$ capture the leading-twist longitudinal tensor polarization structure. The function $b_1$ can be extracted using unpolarized leptons and a spin-1 hadron polarized along the beam (and subtracting the unpolarized contribution):
\begin{equation}
b_1(x)=\frac{1}{2} \left(q^0(x)-q^1(x)\right),
\end{equation}
where 
\begin{equation}
q^0(x)= \left(q_\uparrow^0+q_\downarrow^0\right) = 2q_\uparrow^0, \quad
q^1(x)=  \left(q_\uparrow^1+q_\downarrow^1\right) = \left(q_\uparrow^1+q_\uparrow^{-1}\right). 
\end{equation}
The longitudinal tensor polarization state $S_{LL}$ is also referred to as the "alignment" (see \cite{Bacchetta:2000jk} for instructive pictures of the various polarization states, where the notation $f_{1LL}=b_1$ is used). 
In the parton model one finds the relation $b_2=2xb_1$. The structure function $b_1$ has been measured by the HERMES Collaboration \cite{Airapetian:2005cb} showing it is clearly nonzero in the range $x=0.01-0.1$, but zero below the percent level for higher $x$. A more precise measurement would be interesting, as it is related to the partonic structure of the combined system of a proton and a neutron, not present in the individual nucleons. That it is nonzero below $x=0.1$ may be surprising given that the deuteron is only a loosely bound state. 

A novel measurement possible with a polarized deuteron would be of the transverse tensor polarization ($S_{TT}$ with distribution $h_{1TT}$) that can arise at leading twist but that is absent in the parton model. It thus occurs only due to gluons \cite{Jaffe:1989xy,Artru:1989zv}. 

Due to the gyromagnetic ratio $g$ of the deuteron being 30\% smaller than that of the proton, one finds $G=(g-2)/2=1.79$ for the proton, while $G=-0.14$ for the deuteron, yielding  much lower $|G\gamma|$ values for the deuteron, thus making it harder to spin polarize a beam of deuterons.   
  
\section{3D spin structure}
Transverse momentum dependent parton distributions (TMDs) provide information about the three-dimensional momentum structure $(x,\bm{k}_T)$ and can be probed in not entirely inclusive processes. For example, quark TMDs can be probed in semi-inclusive DIS or SIDIS ($ep \to e' h X$, where the final state hadron $h$ is in the current fragmentation region), and gluon TMDs can be probed in $D$-meson pair production ($ep\to e' D \bar{D} X$, where one has to measure the transverse momentum of the pair). Since the transverse momentum dependence can be correlated with the spin, there are more TMDs than collinear parton distributions. An example is the quark Sivers effect, which at the EIC can be measured in for instance $ep \to e'\, {\rm jet} \, X$ ("jet SIDIS"): 
\begin{equation}
\frac{d\sigma(e \, {p}^\uparrow \to e' \, {\rm jet} \, X)}{d^{2}\bm{q}_T } \propto\;|\bm S_{T}^{}|
\;\sin(\phi_{\small {\rm jet}}^e-\phi_{S}^e)\; \frac{Q_T}{M} {f_{1T}^{\perp q}(x,Q_T^2)}, \qquad
Q_T^2 = |\bm P^{{\rm jet}}_{\perp}|^2.
\end{equation}
One can probe the transverse momentum dependence of the Sivers function $f_{1T}^{\perp q}$ directly in this way. Another advantage of the EIC is that its  measurement is possible in the same kinematic region as in Drell-Yan (DY).  This is important for a clean test of the predicted sign change relation \cite{Collins:2002kn}, $f_{1T}^{\perp q [{\rm SIDIS}]}(x,k_T^2) = - f_{1T}^{\perp q [{\rm DY}]}(x,k_T^2)$, which arises from the difference of initial versus final state interactions (ISI/FSI). Summation of all gluon rescatterings leads to path-ordered exponentials in the TMD correlators, where the integration path depends on the process. FSI lead to a future pointing Wilson line (the $+$ link), whereas ISI to past pointing (the $-$ link). This leads to observable effects, such as a nonzero Sivers asymmetry \cite{Brodsky:2002cx,Brodsky:2002rv}. The Sivers effect in SIDIS has been clearly observed by HERMES at DESY \cite{Airapetian:2009ae} and COMPASS at CERN \cite{Alekseev:2010rw}. The corresponding DY experiments are investigated at CERN (COMPASS), Fermilab (SeaQuest), RHIC (W-boson production) and planned at NICA (Dubna) and IHEP (Protvino). The first data \cite{Adamczyk:2015gyk,Aghasyan:2017jop} are compatible with the sign-change prediction of the TMD formalism. A similar sign change relation holds for gluon Sivers functions \cite{Boer:2016fqd}:
$f_{1T}^{\perp\, g \, [e\, p^\uparrow \to e' \, Q\, \overline{Q}\, X]}(x,p_T^2) = - f_{1T}^{\perp\, g \, [p^\uparrow\,  p\to \gamma \, \gamma \, X]} (x,p_T^2)$. The process on the right hand side could be measured at RHIC, but is very challenging. The left hand side could be measured at EIC but is also quite challenging. The Sivers asymmetry in open heavy quark production $e\, p^\uparrow \to e' \, Q \overline{Q}\, X$ is given by \cite{Boer:2016fqd}:
\begin{equation}
A^{\sin(\phi_S-\phi_{\scriptscriptstyle T})} =  \frac{\vert \bm q_{\scriptscriptstyle T}\vert}{M_p}\, \frac{f_{1T}^{\perp\,g}(x,\bm q_{\scriptscriptstyle T}^2)}{f_1^g(x,\bm q_{\scriptscriptstyle T}^2)}.
\end{equation}
The maximally allowed gluon Sivers function would give 1 for this asymmetry. However, if the function is 10\% of this bound, then assuming $L_{\rm int} = 10\ {\rm fb}^{-1}$ it cannot be discerned within the statistics \cite{Zheng:2018awe}. 
The situation for dijets is more promising, but that is theoretically less clean \cite{Boer:2016fqd}.

Unpolarized open heavy quark pair production in $ep$ (and $eA$) collisions offers another interesting opportunity:
to probe linearly polarized gluons in unpolarized hadrons. It gives rise to an angular distribution: a $\cos 2(\phi_{\scriptscriptstyle T} - \phi_\perp)$ asymmetry, where $\phi_{\scriptscriptstyle T} - \phi_\perp$ are the angles of the sum and difference transverse momentum of the two heavy quarks.  In this observable the distribution of linearly polarized gluons $h_1^{\perp\; g}$ appears by itself, so effects could be significant, especially towards smaller $x$. It is expected to keep up with the growth of the unpolarized gluons as $x \to 0$. The maximally allowed asymmetries are substantial (for any $Q^2$ and for both charm and bottom) \cite{Pisano:2013cya}. A small-$x$ model (the MV model that is expected to be relevant at $x$ around 0.01) gives similar results \cite{Boer:2016fqd}. Apart from heavy quark pair production, $h_1^{\perp\; g}$ is accessible in dijet production at a high-energy EIC \cite{Metz:2011wb,Pisano:2013cya,Boer:2016fqd}. Also here the gluon polarization shows itself through a $\cos 2\phi$ distribution. A large azimuthal modulation is expected \cite{Dumitru:2015gaa}. Interestingly, the $\cos 2\phi$ modulation has opposite signs for $L$ and $T$ polarization states of the virtual photon \cite{Dumitru:2018kuw}.
  
Another process at EIC that can be used to probe polarized TMDs is quarkonium production: $e \, p^\uparrow \to e' \, {[Q\overline{Q}]} \, X$, where the quarkonium bound state $[Q\overline{Q}]$ can be a $J/\psi$ or $\Upsilon$ state \cite{Godbole:2012bx,Godbole:2013bca,Godbole:2014tha,Mukherjee:2016qxa,Rajesh:2018qks}. Since in leading order (LO) the quarkonium is in a color octet state, one either considers the Color Evaporation Model or NRQCD for Color Octet (CO) states. In the latter approach the spin asymmetries depend on the quite uncertain CO NRQCD long distance matrix elements (LDMEs), but one can consider ratios of asymmetries to cancel those out \cite{Bacchetta:2018ivt}. Conversely, one can consider ratios where the TMDs cancel out (in leading order) and one can obtain new experimental information on the CO NRQCD LDMEs. This requires a comparison to the process of open heavy quark pair production $e \, p \to e' \, Q \, \overline{Q} \, X$:
\begin{eqnarray}
&& {\cal R}^{\cos2\phi_{\scriptscriptstyle T}} = \frac{\int  d \phi_{\scriptscriptstyle T} \cos 2\phi_{\scriptscriptstyle T}\, d\sigma^{[Q\overline{Q}]} (\phi_S,\phi_{\scriptscriptstyle T})}{\int  d \phi_{\scriptscriptstyle T} \, d\phi_\perp \cos 2\phi_{\scriptscriptstyle T}\, d\sigma^{Q\overline Q} (\phi_S,\phi_{\scriptscriptstyle T}, \phi_\perp) } \stackrel{{\rm LO}}{=} 
\frac{27 \pi^2}{4}\, \frac{1}{M_Q} \,  \left [{\cal O}_8^S - \frac{1}{M_Q^2} \, {\cal O}_8^P \right ] \,,
\\[2 mm]
&& {\cal R} =  \frac{\int  d \phi_{\scriptscriptstyle T} \, d\sigma^{[Q\overline{Q}]} (\phi_S,\phi_{\scriptscriptstyle T})}{\int  d \phi_{\scriptscriptstyle T} \, d\phi_\perp\, d\sigma^{Q\overline Q} (\phi_S,\phi_{\scriptscriptstyle T}, \phi_\perp) }
\stackrel{{\rm LO}}{=} 
\frac{27\, \pi^2}{4}\,\frac{1}{M_Q} \, \frac{ [ 1+(1-y)^2 ] \, {\cal O}_{8}^S + ( 10-10y+3y^2  )\, {\cal O}_{8}^P /M_{Q}^{2}}{26 -26 y +9 y^2}  \, .\nonumber
\end{eqnarray}
In this way one obtains two observables depending on two "unknowns", the two CO NRQCD LDMEs $ {\cal O}_{8}^S \equiv  \langle0\vert{\cal O}_{8}^{[Q\overline{Q}]}(^{1}S_{0})\vert0\rangle $ and $ {\cal O}_{8}^P \equiv  \langle0\vert{\cal O}_{8}^{[Q\overline{Q}]}(^{3}P_{0})\vert0\rangle $. Two similar, but different observables depending on the same two unknowns are obtained for polarized quarkonium production, offering a way to cross-check the results \cite{Bacchetta:2018ivt}.

\section{Fragmentation functions} 

The process of semi-inclusive DIS with transversely polarized protons, $ep^\uparrow \to e'\, h \, X$, can be exploited to probe the Sivers effect through a $\sin(\phi_S-\phi_h)$ modulation, but it also receives contributions from transversely polarized quarks showing up as a $\sin(\phi_S+\phi_h)$ modulation. This asymmetry is a convolution of the transversity distribution $h_1^q$ and the Collins effect FF. Both are transverse momentum dependent. One can also use two-hadron fragmentation functions to probe quark transversity, in which case one can consider the transverse momentum integrated case \cite{Collins:1993kq,Collins:1994ax,Jaffe:1997hf,Bianconi:1999cd,Radici:2001na,Radici:2018iag}.
The quark helicity distributions $g_1^q=\Delta q$ can also be probed using two-hadron fragmentation functions, but it requires transverse momentum. It enters with the "handedness" fragmentation functions $G_1^\perp$ \cite{Bianconi:1999cd}, which according to a 
model calculation \cite{Matevosyan:2018oui} could be of order 10\% of the unpolarized FF $D_1$. It can be extracted from Belle data \cite{Abdesselam:2015nxn}, but in a rather more involved way than was suggested initially, see \cite{Matevosyan:2017liq,Matevosyan:2018icf} for details.  
  
Polarized $\Lambda$'s can also be used to probe $g_1^q$ via the polarization transfer $D_{LL}$ \cite{Belostotski:2011zza,Alekseev:2009ab,Adam:2018kzl}. Similarly the transversity distribution $h_1^q$ can be extracted from $D_{NN}$ in SIDIS. COMPASS data indicates a small polarization transfer \cite{Negrini:2009oia}. This could be due to the transversity FF $H_1^{u,d}(z)$ and/or the strange quark transversity $h_1^s(x)$ being small in the measured range. 
  
Produced $\Lambda$'s can also become "spontaneously" polarized, as is long known from $\Lambda$ production in $pA$ collisions. In the TMD formalism this can arise from unpolarized quarks or gluons fragmenting into transversely polarized $\Lambda$'s, described by the "polarizing" TMD FF $D_{1T}^\perp$ \cite{Mulders:1995dh}. It could provide the underlying description of $p A \to \Lambda^\uparrow X$ \cite{Anselmino:2000vs}, although it is not a TMD process. Measurements of $\Lambda$ production in SIDIS could help clarify the mechanism \cite{Anselmino:2001js}. There is data on $\Lambda$ production in charged and neutral current DIS but mostly these are in the target fragmentation region or in quasi-real production ($Q^2\approx 0$) \cite{Adams:1999px,Belostotsky:1999kj}. The only available SIDIS data in the current fragmentation region is from NOMAD ($\nu_\mu p\to \mu \Lambda^\uparrow X$) \cite{Astier:2000ax} and from ZEUS ($ep\to e\Lambda^\uparrow X$) \cite{Chekanov:2006wz}, which are both compatible with zero within large errors. 
  
Polarizing FFs can be extracted from $e^+ e^-$ experiments. It was pointed out in \cite{Boer:1997mf} that 
in $e^+ e^ - \to \Lambda^\uparrow \, {\rm jet}\ X$, where the $\Lambda$ is part of the opposite side jet, its contribution is not power suppressed, unlike in $e^+ e^ - \to \Lambda^\uparrow \ X$, e.g.\ considered in \cite{Gamberg:2018fwy}.
Data from LEP ($Q=M_Z$) on $e^+ e^ - \to \Lambda^\uparrow \, {\rm jet}\ X$, where the jet axis is approximated by the thrust axis, is compatible with zero at the $\sim 3\%$ level \cite{Ackerstaff:1997nh}, but recent data by the BELLE Collaboration ($Q=10.58$ GeV) \cite{Abdesselam:2016nym,Guan:2018ckx} is clearly nonzero and ranges up to the 10\% level as a function of $p_T$ of the $\Lambda$ w.r.t.\ the thrust axis. This is compatible with the expectation from TMD evolution \cite{Boer:2001he} that the polarization depends on $Q$ roughly as $Q^{-0.6}$, implying that the polarization at BELLE is expected to be about 3 times larger than at $Q=M_Z$.  
  
As pointed out in \cite{Boer:2010ya} $e^+ e^ - \to \Lambda \, {\rm jet}\ X$ is very sensitive to cancellations between $u$, $d$ and $s$ contributions. It is better to study $\Lambda$'s produced in association with an opposite side charged pion or kaon, $e^+ e^- \to \Lambda \, h^\pm \ X$, which allows for flavor selection. Comparison to $ep \to e \Lambda^\uparrow X$ can be used to test the expected universality of $D_{1T}^\perp$.  BELLE data on $e^+ e^- \to \Lambda \, h^\pm \ X$ \cite{Abdesselam:2016nym,Guan:2018ckx} does not follow the expectations of \cite{Boer:2010ya} for the hadron charges, but those predictions were based on the still rather poorly known unpolarized $u,d,s \to \Lambda$ FFs.  This discrepancy needs to be looked into (preferably in combination with the still unsolved discrepancy for the gluon to $\Lambda$ FF obtained from fits of $e^+e^-$ data with and without including $pp$ data \cite{Albino:2008fy}).

\section{GTMDs \& GPDs}

Generalized TMDs (GTMDs) are off-forward TMDs, where the off-forwardness of the matrix element is given by $\Delta=p'-p$. GTMDs are Fourier transforms of Wigner distributions \cite{Ji:2003ak,Belitsky:2003nz,Meissner:2009ww}:
\begin{equation}
G(x,\bm{k}_T,\bm{\Delta}_T) \stackrel{{\rm F.T.}}{\longleftrightarrow} W(x,\bm{k}_T,\bm{b}_T)
\end{equation} 
Quark Wigner distributions can display distortions in the $b_T$ plane depending 
on $k_T$ and vice versa, which vanish upon $b_T$ or $k_T$ integration \cite{Lorce:2011kd}.  
Such $b_T \times k_T$ distortions are generally spin-dependent and related to quark orbital angular momentum \cite{Lorce:2011kd}. 

Analogously, gluon Wigner distributions and gluon GTMDs can be defined.
See for a recent review \cite{More:2017zqp}. For an unpolarized proton there are four gluon GTMDs at leading twist. The first suggestion to measure gluon GTMDs is through the process of hard diffractive dijet production \cite{Altinoluk:2015dpi,Hatta:2016dxp}, which extends an earlier suggestion 
to probe gluon GPDs through this process \cite{Braun:2005rg}. This process probes dipole gluon GTMDs (those with one $+$ and one $-$ link), which at leading twist reduce to just one gluon GTMD in the limit $x\to0$ \cite{Boer:2018vdi}. Although any GTMDs can have an "elliptic" part $\propto \cos 2 \phi_{\bm{k} \bm{\Delta}}$, the so-called "elliptic" gluon GTMD from the literature \cite{Hatta:2016dxp,Zhou:2016rnt} refers to the elliptic part of that single small-$x$ dipole gluon GTMD.   
The small-$x$ description of DVCS requires inclusion of the corresponding elliptic Wigner function. It contributes to the helicity flip or transversity gluon GPD $E_T$ \cite{Hatta:2017cte}.
  
At EIC quark GPDs (i.e.\ $k_T$-integrated GTMDs) will be extracted in order to study quark orbital angular momentum as it enters the spin sum rule. Sivers-like distortions ($b_T \times S_T$) and transversity GPDs can also be studied via transverse spin asymmetries. For more information on these investigations see \cite{Boer:2011fh,Accardi:2012qut}. 
    
\section{Conclusions}
The spin physics program at EIC is extremely rich. It includes electroweak polarized structure functions, numerous quark and gluon TMDs, GTMDs and GPDs. Polarized deuterons and neutrons offer further opportunities to learn about the spin of quarks and gluons inside hadrons. Many possible final states allow to probe particular spin effects. Heavy quarks (both open and bound) could prove very useful analyzers of gluon TMDs but also of color-octet NRQCD long distance matrix elements. Polarization dependent fragmentation functions for $\Lambda$'s and hadron pairs offer further tools, but are also interesting in themselves, as there can be polarization transfer effects in the fragmentation process. All these options have considerable interplay and synergy with $e^+e^-$ and (polarized and unpolarized) $pp$ collisions. Spin physics effects at higher twist or their nuclear dependence were not addressed in this overview. The EIC will be essential for small-$x$ and for high-$Q^2$ spin structure studies.

\end{document}